\documentstyle[prd,aps,preprint,floats,epsfig]{revtex}
\tightenlines
\newcommand{\beq}{\begin{equation}}
\newcommand{\eeq}{\end{equation}}
\newcommand{\beqr}{\begin{displaymath}}
\newcommand{\eeqr}{\end{displaymath}}
\newcommand{\beqa}{\begin{eqnarray}}
\newcommand{\eeqa}{\end{eqnarray}}
\newcommand{\beqar}{\begin{eqnarray*}}
\newcommand{\eeqar}{\end{eqnarray*}}

\newcommand{\ssc}{\scriptscriptstyle}
\newcommand{\eg}{{\it e.g.}}
\newcommand{\ie}{{\it i.e.}}

\begin{document}

\thispagestyle{empty}
\rightline{\hfill hep-ph/0104273}
\vskip -.1ex
\rightline{\hfill McGill/01-05}
\vspace*{2cm}

\begin{center}
{\bf \Large Geometry of Large Extra Dimensions versus Graviton Emission}
\vspace*{1cm}

{\bf Fr\'ed\'eric Leblond}\footnote{E-mail: fleblond@hep.physics.mcgill.ca} 
\vspace*{0.3cm}

{\it Department of Physics, McGill University}\\
{\it 3600 University Street, Montr\'eal, Qu\'ebec, H3A 2T8, Canada}\\
\vspace{2cm}
ABSTRACT
\end{center}
We study how the geometry of large extra dimensions may affect field theory results
on a three-brane. More specifically, we compare cross sections for graviton emission from a brane when the
internal space is a $N$-torus and a $N$-sphere for $N=2$ to 6. The method we present can be used
for other smooth compact geometries. 
We find that the ability of high energy colliders to determine the geometry of the extra dimensions
is limited but there is an enhancement when both the quantum gravity scale and $N$ are large.
Our field theory results are compared with the low energy corrections to the
gravitational inverse square law due to large dimensions compactified on other spaces such as Calabi-Yau manifolds.  

\vfill
\setcounter{page}{0}
\setcounter{footnote}{0}
\newpage

\section{Introduction}
In recent years we have learned there could be more than meets the eye concerning gravity. While this is expected what is rather
surprising is that we can appreciate this statement without entering the realm of M-theory.
The idea that there might be more than the commonsensical four dimensions of our everyday world has been floating around for
many decades now. This very fruitful idea was used in many contexts sometimes, it seems, according to taste. Its most fundamental incarnation
is found in string/M-theory where the extra dimensions are introduced for consistency. Which vacuum this elegant unifying scheme
ultimately chooses is an open problem. Consequently, there are many possibilities as to what the low energy theory of gravitation
in our universe can be. For example, effective theories with a factorizable metric and a 
Planck scale of energy lower than the one associated with the four-dimensional gravitational 
coupling ($M_{\ssc{P}}\approx 1.30\times 10^{19}$ {\rm GeV}) are not excluded \cite{ADD}. In fact, there is a vast 
literature on the idea of using extra dimensions and a lower quantum gravity scale to devise effective gravitational models. 
Inspired by the $Dp$-brane concept of string theory, the Standard Model (SM) fields are assumed to be localized on a three-brane
(the classical version of a stack of $D3$-brane) while gravity propagates in the entire spacetime. 
An interesting feature of this scenario is the potentially large size of the extra dimensions. For example, a brane world model
with $N=2$ transverse dimensions and a quantum gravity scale, $M_{\ssc{D}}$, of the order of 1 {\rm TeV} leads to dimensions that
can be as large as one millimeter. Although this particular set of parameters seems to be ruled out by astrophysical bounds
\cite{astro}, it is still worth investigating the large dimensions scenario for other values of $N$ and $M_{\ssc{D}}$.   

When the effects of large extra dimensions on SM processes are studied these are usually compactified on 
a $N$-dimensional torus \cite{wells,hewett,peskin,rizzo}. In this work, we carefully study some of the consequences of having the 
extra dimensions compactified 
on a $N$-sphere. In sections \ref{gravity} and \ref{spherical} we describe the linear theory of gravity on which we build 
our work and
comment on the Kaluza-Klein compactification scheme we use. In the following section we perform the mode decomposition of
the graviton both for toric and spherical internal manifolds. In section \ref{emission} we devise a method for computing
cross sections when large dimensions are compactified on smooth geometries. Explicit calculations are performed
in order to compare models with spaces compactified on a torus and a sphere. We find that gravitons having a small 
momentum (transverse to the brane) with respect to the curvature of the extra dimensions are the most efficient for studying the geometry of the
compact space. This leads to the observation that high energy colliders have limited power in determining the geometry
of the extra dimensions since the relative importance of the low energy modes becomes less important for processes 
with large $\sqrt{s}$. We also compare our results with the low energy potential obtained from gravitational 
models with large extra dimensions compactified
on tori, spheres and some Calabi-Yau manifolds. Finally we comment on possible extensions of our work leading to effects
potentially detectable at the {\rm LHC}.
               
\section{Linear Gravity and Kaluza-Klein Compactification}
\label{gravity}
The degrees of freedom we consider are those of general relativity 
which we use as our effective theory. Of course, the physics at high energy will leave residual effects at low energy but these are
highly suppressed non-renormalizable interactions. The quantized ($4+N$)-dimensional version of gravity we use follows 
the covariant approach (for example see Ref.~\cite{wheeler}) in which the graviton field, $h_{\ssc{AB}}$, is a small perturbation,
\beq
\label{RFP}
g_{\ssc{AB}} = \bar{g}_{\ssc{AB}} + h_{\ssc{AB}} \;\;\;\;\;\;\; |h_{\ssc{AB}}| \ll 1 \; , 
\eeq          
where $A,B = 0,1,...(D-1)$. The result is a perturbative theory on a background with metric
$\bar{g}_{\ssc{AB}}$. We are considering quantization around $M^{\ssc{4}}\times T^{\ssc{N}}$ 
and $M^{\ssc{4}}\times S^{\ssc{N}}$ where $M^{\ssc{4}}$ is four-dimensional Minkowski space. This work is therefore based on the linear
version of the Einstein equations with the following metric ansatz 
\begin{displaymath}
g_{\ssc{AB}} = \left( \begin{array}{cc} \eta_{\mu\nu} & 0 \\ 0 & g_{ij} \end{array} \right) + h_{\ssc{AB}},
\end{displaymath}
where $g_{ij}$ ($i=1$ to $N$) is the metric of the large transverse dimensions (manifold $B^{\ssc{N}}$) and $\mu = 0,1,2,3$. 
We denote coordinates on $M^{\ssc{4}}$ (the three-brane) with $x^{\mu}$ and those in the internal space with $y^{i}$.
It is understood that a model with non-trivial $g_{ij}$ might not be a solution of the field equations 
without the addition of matter to the system. We address some such issues in Ref. \cite{leblond2}. In Sec.~\ref{spherical},
we show what ingredients are needed to build a model with a flat brane and the extra dimensions compactified
on a $N$-sphere in a manner which is consistent with the field equations.   
   
Once quantized, the states of this theory are $D$-dimensional spin-two plane waves. The physical states are the gauge 
fixed ones. Picking the harmonic gauge condition takes out $(4+N)$ degrees of freedom but there is a residual set 
of diffeomorphisms preserving this gauge choice
$\ie$
\beq
\label{GCTkk}
x_{\ssc{A}}\rightarrow x_{\ssc{A}} + \xi_{\ssc{A}},
\eeq  
with $\Box_{\ssc{D}}\xi_{\ssc{A}}=0$ ($\Box_{\ssc{D}}$ is the $D$-dimensional d'Alembertian operator on
$M^{\ssc{4}}\times B^{\ssc{N}}$). Once all the gauge freedom is used up, we are left with physical ($4+N$)-dimensional
gravitons having 
\beq
\frac{(N+2)(N+3)}{2}-1
\eeq
polarization states.

We now explain how small perturbations around $\bar{g}_{\ssc{AB}}$ manifest themselves from a four-dimensional point 
of view. Since we assume
$B^{\ssc{N}}$ to be compact the correct procedure is to perform a Kaluza-Klein decomposition of 
$h_{\ssc{AB}}$ (see for example Refs.~\cite{appel}). Symmetric spaces
are characterized by Killing vectors. The set of such vectors for a manifold with Euclidean signature represents 
the family of one-parameter diffeomorphic transformations leaving the metric invariant,   
\beq
\label{lie}
\pounds_{K^{a}} g_{ij} = \nabla_{(i} K^{a}_{j)}=0, 
\eeq
where $\pounds_{K^{a}}$ is the Lie derivative along $K^{a}$.
A maximally symmetric $N$-dimensional manifold has $N(N+1)/2$ Killing vectors ($\eg$ $T^{\ssc{N}}$, 
$S^{\ssc{N}}$). The $K^{a}$'s 
obey the Lie algebra of a group $G$ that depends on the isometries of $B^{\ssc{N}}$,
\beq
\label{lie}
[K^{a},K^{b}] = f^{\;ab}_{c} K^{c}.
\eeq 
For example, $S^{\ssc{2}}$ has three Killing vectors transforming under $SO(3)$.

A special class of coordinate transformations on $B^{\ssc{N}}$ is the isometry group
\beq
\label{translations}
y^{i} \rightarrow y^{i} + \epsilon^{a}(x)K^{i}_{a}(y) 
\eeq
$\ie$ the family of diffeomorphisms defined on the compact geometry that leave the metric unchanged. 
Eq.~(\ref{lie}) shows that it is natural to associate a group $G$ with $B^{\ssc{N}}$ ($\eg$ $SU(2)$ for $S^{\ssc{2}}$). 
Then to 
each of the Killing
vectors we associate a group generator $T^{a}$. These can be thought of as generating the symmetries (\ref{translations}) of
$G$ on $B^{\ssc{N}}$. Each $T^{a}$ corresponds to a gauge boson $A_{\mu}^{a}$. We write the following general ansatz for the massless
modes of $h_{\ssc{AB}}$,
\beq
\label{KKan}
h_{\ssc{AB}} = f(V_{\ssc{N}})\left( \begin{array}{cc} h_{\mu\nu} + \eta_{\mu\nu}\phi_{ii} & A^{a}_{\nu}K^{a}_{i} \\ 
A^{a}_{\mu}K^{a}_{i} & -N \phi_{ij} \end{array} \right),
\eeq
where $\phi_{ij}$ is a matrix of scalar fields and $f(V_{\ssc{N}})$ is a function of the compact space volume.
The diffeomorphic transformations (\ref{translations}) are equivalent to
\beq
\label{covariant}
A_{\mu}^{a} \rightarrow A_{\mu}^{a} + D_{\mu}\epsilon^{a},
\eeq
where $D_{\mu}$ is the covariant derivative associated with the resulting fiber bundle on $M^{4}$. The transformation rule
(\ref{covariant}) could, for example, correspond to the gauge transformations of a non-abelian theory. By varying the 
topology of $B^{\ssc{N}}$ we can obtain different gauge theories. So $G$ is a subgroup of the $(4+N)$-dimensional 
diffeomorphisms 
perceived as an internal symmetry from a four-dimensional point of view. The simplest case is when 
$B^{\ssc{N}}=T^{\ssc{N}}$. Then the Kaluza-Klein
ansatz for the metric perturbation takes the following form,
\beq
\label{KKtorus}
h_{\ssc{AB}} = \frac{1}{\sqrt{V_{T^{\ssc{N}}}}}\left( \begin{array}{cc} h_{\mu\nu} + \eta_{\mu\nu}\phi_{ii} & A_{\mu i} \\ 
A_{\mu i} & -N \phi_{ij} \end{array} \right),
\eeq
where $V_{T^{\ssc{N}}}$ is the volume of a $N$-torus and $A_{\mu i}$ are abelian gauge bosons.

Ultimately, we want to compare cross sections for processes taking place on $M^{\ssc{4}}$ when the topology of $B^{\ssc{N}}$ 
is modified.
For the kind of processes we are investigating only the spin-two part of 
the metric perturbation, $h_{\mu\nu}$, is relevant. 

\section{Field equations of the spin-two perturbation when the extra dimensions are compactified on a $N$-sphere}
\label{spherical}
Our aim is to compare cross sections for processes in models with extra dimensions compactified on spheres and tori.
To accomplish this we use models containing flat three-branes. In a scenario with toric manifolds this is easily
realized since the curvature in the internal space is zero. This is not the case when the geometry of the extra dimensions 
is spherical. Then, the curvature of the extra dimensions
forces us to modify the model by, for example, introducing a bulk cosmological constant and an abelian gauge field trapped inside
the compact manifold. Only then can we obtain a model with a flat brane that is consistent with the Einstein equations. 
The field equations are then
\beq
R_{\ssc{AB}} - \frac{1}{2}g_{\ssc{AB}} R = - 8\pi G_{\ssc{D}}\,\frac{1}{\sqrt{-g}} 
\frac{\delta S_{\ssc{M}}}{\delta g^{\ssc{AB}}},
\eeq
where $S_{\ssc{M}}$ is the action for a bulk cosmological constant, $\Lambda$, and a $p$-form gauge field, $A_{p}$,
with field strength $F_{p+1}=d A_{p}$,
\beq
S_{\ssc{M}} = \int d^{4}x d^{\ssc{N}}y \sqrt{-g} \left[ -2\Lambda -\frac{1}{2(p+1)!}F_{\ssc{A_{1}}...\ssc{A_{p+1}}}
F^{\ssc{A_{1}}...\ssc{A_{p+1}}} \right].
\eeq
The resulting field equations are
\beq
R_{\ssc{AB}} - \frac{1}{2}g_{\ssc{AB}} R = -\Lambda g_{\ssc{AB}} -\frac{4\pi G_{\ssc{D}}}{(p+1)!} \left[ \frac{1}{2}g_{\ssc{AB}}
F_{\ssc{A_{1}}...\ssc{A_{p+1}}} F^{\ssc{A_{1}}...\ssc{A_{p+1}}} - (p+1) F_{\ssc{A}}{}^{\ssc{A_{2}}...\ssc{A_{p+1}}}
F_{\ssc{B}\ssc{A_{2}}...\ssc{A_{p+1}}} \right].
\eeq 
We now specialize to a model with a flat tangent space metric $\eta_{\mu\nu}$, a tranverse space with the metric $g_{ij}$
of a $N$-sphere and a $(N-1)$-form (magnetic) field trapped in the compact space. The latter is accomplished by using
the ansatz $F_{\ssc{A_{1}}...\ssc{A_{N}}} = k \; \delta^{i_{1}}_{\ssc{A_{1}}}...\delta^{i_{N}}_{\ssc{A_{N}}}
\epsilon_{i_{i}...i_{p+1}}$ where $k$ is a constant. The Einstein equations then reduce to
\beq
\label{equationE1}
-\frac{1}{2}\eta_{\mu\nu} \, \tilde{R} = [-\Lambda - 2\pi G_{\ssc{D}} k^{2}] \,\eta_{\mu\nu}, 
\eeq 
\beq
\label{equationE2}
R_{ij} -\frac{1}{2} g_{ij} \, \tilde{R} = [-\Lambda + 2\pi G_{\ssc{D}} k^{2}] \, g_{ij}, 
\eeq
where $\tilde{R} = N(N-1)/a^{2}$ is the Ricci scalar in the internal space and
\beq
\label{ricciextra}
R_{ij} -\frac{1}{2} g_{ij} \, \tilde{R} = -\frac{1}{2 a^{2}} (N-2)(N-1) \, g_{ij},
\eeq
with $a$ the radius of the spheres. Using the expression for $\tilde{R}$ and Eq.~(\ref{ricciextra}),
Eqs.~(\ref{equationE1}) and (\ref{equationE2}) reduce to
\beq
\Lambda + 2\pi G_{\ssc{D}} k^{2} = \frac{N(N-1)}{2 a^{2}},
\eeq   
\beq
\Lambda - 2\pi G_{\ssc{D}} k^{2} = \frac{1}{2 a^{2}}(N-2)(N-1).
\eeq
Inspection of the last two equations shows why it is necessary to introduce both a bulk cosmological constant
and a gauge field for a model with a flat three-brane to satisfy the gravitational field equations when the extra dimensions
are compactified on a $N$-sphere. 

We now derive the linear free field equations that the spin-two perturbation, $h_{\mu\nu}$, must satisfy when propagating in the
background $M^{4}\times S^{\ssc{N}}$ described above. In order to achieve this we use the simplified Kaluza-Klein ansatz
\beq
\label{simpleansatz}
h_{\ssc{AB}}(x,{\bf y}) = \frac{1}{\sqrt{V_{S^{\ssc{N}}}}}\left( \begin{array}{cc} h_{\mu\nu}(x,{\bf y}) & 0 \\ 
0 & 0 \end{array} \right).
\eeq
In the linear limit, the free field equations with both the cosmological constant and the gauge field turned on are
\beq
R_{\mu\nu}^{\ssc{(L_{1})}} + R_{\mu\nu}^{\ssc{(L_{2})}} = 0 
\eeq
and
\beq
\label{constrainteom}
R^{\ssc{(L)}}_{ij} = -\frac{1}{2}\partial_{i}\partial_{j} h = 0,
\eeq
where $h$ is the trace of $h_{\mu\nu}$ and
\beq
\label{compli1}
R^{\ssc{(L_{1})}}_{\mu\nu}= -\frac{1}{2} \partial^{\lambda}\partial_{\lambda} h_{\mu\nu} + \frac{1}{2} [\partial_{\mu}(\partial_{\lambda}h^{\lambda}_{\nu}
-\frac{1}{2}\partial_{\nu}h) + \partial_{\nu}(\partial_{\lambda}h^{\lambda}_{\mu}
-\frac{1}{2}\partial_{\mu}h)],
\eeq
\beq
\label{compli2}
R^{\ssc{(L_{2})}}_{\mu\nu}= -\frac{1}{2} \partial^{i}\partial_{i} h_{\mu\nu} - \frac{1}{2}[\partial_{j}h_{\mu\nu} \partial_{i}g^{ij} + \Gamma^{i}_{ij}
g^{jk}\partial_{k}h_{\mu\nu}].
\eeq
Eq.~(\ref{constrainteom}) is a constraint on the trace, $h$, of the spin-two field which is present even when the
internal space is flat. It constrains all Kaluza-Klein modes of the graviton to be traceless (see Sec.~\ref{graviton} for more
details on the Kaluza-Klein decomposition procedure). 
To simplify Eq.~(\ref{compli1}) we use the harmonic gauge condition $g^{\mu\nu}\Gamma^{\ssc{C}}_
{\mu\nu} = 0$ which is equivalent to
\beq
\label{donder1}
\partial_{\mu}h^{\mu}_{\lambda} -\frac{1}{2}\partial_{\lambda} h = 0.
\eeq
Using Eq.~(\ref{donder1}) the tensor $R^{\ssc{(L_{1})}}_{\mu\nu}$ simplifies to 
\beq
\label{part1}
R^{\ssc{(L_{1})}}_{\mu\nu} = -\frac{1}{2}\partial^{\lambda}\partial_{\lambda} h_{\mu\nu}. 
\eeq
Now from $\nabla_{i}g^{ij} = 0$ we find
\beq
\label{simpli}
\partial_{i}g^{ij} = -\Gamma^{i}_{ik} g^{kj} - \Gamma^{j}_{ik} g^{ik},
\eeq
which allows us to write 
\beq
\label{part2}
R^{(2)}_{\mu\nu} = -\frac{1}{2}\nabla^{i}\nabla_{i} h_{\mu\nu}, 
\eeq
where use has also been made of the fact that, when applied on a scalar function, the following operator equality holds, 
\beq
g^{ij}\nabla_{i}\nabla_{j} = g^{ij}\partial_{i} \partial_{j} - g^{ij} \Gamma^{k}_{ij}\partial_{k}. 
\eeq
Combining Eqs.~(\ref{part1}) and (\ref{part2}) the linear free field equations for the spin-two field $h_{\mu\nu}$ become
\beq
\label{KG}
\Box_{\ssc{D}} \, h_{\mu\nu}(x,{\bf y}) = 0,
\eeq
where $\Box_{\ssc{D}}$ is the d'Alembertian operator on $M^{\ssc{4}}\times S^{\ssc{N}}$. Similarly, the $(i \, j)$ 
components of 
pertubation (\ref{KKan}), when the $(\mu \, \nu)$ and $(\mu \, i)$ components are turned off, satisfy an equation 
similar to Eq.~(\ref{KG}). Moreover, the off-diagonal $(\mu \, i)$ components satisfy Maxwell-like
field equations when the $(\mu \, \nu)$ and $(i \, j)$ components are ignored. When all components of the perturbation are turned on, the physical fields correspond to mixings
between the metric components. The full calculation is presented in Ref.~\cite{lykken} for the case of large extra dimensions
compactified on a torus.  

\section{Graviton Modes in the Internal Space}
\label{graviton}

A detailed analysis of the Kaluza-Klein reduction and gauge fixing procedure for compactification on a $N$-torus can be found in 
Ref.~\cite{lykken}. Using a similar procedure one expects to find the following ($4+N$)-dimensional equations of motion for the
spin-two perturbation (which we call from now on the graviton):
\beq
\label{boxh}
\Box_{\ssc{D}} \, h_{\mu\nu}(x,{\bf y}) = 0,
\eeq
where $\Box_{\ssc{D}}$ is now the d'Alembertian operator on $M^{\ssc{4}}\times B^{\ssc{N}}$ (although we have demonstrated this to be true 
explicitely only for $B^{\ssc{N}}=S^{\ssc{N}}$).   
Because of the compact nature of $B^{\ssc{N}}$ the gravitational field can be recast as an infinite tower of Kaluza-Klein (KK) 
modes,
\beq
\label{eomg}
h_{\mu\nu}(x,{\bf y})= \sum_{\{n\}} h_{\mu\nu}^{\{n\}}(x) \psi^{\{n\}}({\bf y}),
\eeq
where $\{n\}$ is a set of quantum numbers related to the isometry group of the compact space and $\psi^{\{n\}}({\bf y})$ is
a normalized wave function. As can be read from Eq.~(\ref{boxh}) the wave functions
must be such that
\beq
\label{wave}
\left[ \nabla_{i}\nabla^{i} + m_{\{n\}}^{2} \right] \psi^{\{n\}}({\bf y}) = 0,
\eeq
which we solve both for the $T^{\ssc{N}}$ and $S^{\ssc{N}}$ geometries. We see that the KK modes are the eigenmodes of 
the appropriate Laplacian
on the internal space. They depend completely on its geometry and topology. Compact hyperbolic spaces have been considered
in Ref.~\cite{hyper}.

\subsection{Compactification on a $N$-Torus}
The simplest compact geometry for the extra dimensions is a $N$-dimensional torus 
with a unique radius $a$. Cases with toric extra dimensions characterized by different length scales are studied in 
Ref.~\cite{lykken2}. The wave 
function in transverse space is simply obtained by solving
\beq
\left[ \partial_{i}\partial^{i}+m^{2}_{{\bf n}} \right] \psi^{{\bf n}}({\bf y}) = 0,
\eeq
which leads to
\beq
\label{Ntorus}
\psi^{{\bf n}}({\bf y}) = \frac{1}{(2\pi a)^{\ssc{\frac{N}{2}}}} e^{i\ssc{\frac{{\bf n}\cdot {\bf y}}{a}}},
\eeq 
where ${\bf n}=\{n_{1},n_{2},...,n_{\ssc{N}}\}$ with the $n_{i}$'s integers running from $-\infty$ to $+\infty$ and 
$0<y_{i}\leq 2\pi a$ are the components of the vector ${\bf y}$. Based on Eq.~(\ref{KKtorus}) the ${\bf n}={\bf 0}$ modes correspond to a massless graviton
($2$ degrees of freedom), a set of $N$ $U(1)$ massless gauge bosons ($2N$ degrees of freedom) and moduli composed of 
$N(N+1)/2$
massless scalars for an expected total of $(2+N)(3+N)/2 - 1$ degrees of freedom. For ${\bf n}\neq {\bf 0}$ the analysis is somewhat
different since then
the momentum of the graviton in the transverse space is not zero. 
For an observer on the brane this transverse momentum is perceived as a  
four-dimensional mass ($m^{2}={\bf n}^{2}/a^{2}$). The spectrum then consists   
(for each level ${\bf n}\neq {\bf 0}$) of one massive spin-two particle, $(N-1)$ massive vector bosons and a set of
$N(N-1)/2$ massive
scalars. For later reference we write down the wave function at ${\bf y}={\bf 0}$,
\beq
\label{toricw}
\psi^{{\bf n}}({\bf 0}) = \frac{1}{(2\pi a)^{\ssc{\frac{N}{2}}}} = \frac{1}{\sqrt{V_{\ssc{T^{N}}}}},
\eeq  
where $V_{\ssc{T^{N}}}$ is the volume of the torus.

\subsection{Compactification on a $N$-Sphere}
\label{psi}
The derivation of the wave function for the graviton propagating on a $N$-sphere is more challenging. 
In that case we need to solve the following equation, 
\beq
\label{spherewave}
\left( \nabla^{2}_{\ssc{S^{N}}} + m_{\{n\}}^{2} \right) \psi^{\{n\}}({\bf y}) = 0,
\eeq
where $\nabla^{2}_{\ssc{S^{N}}}$ is the Laplacian on a $N$-sphere of fixed radius $a$,
\beq
\nabla^{2}_{\ssc{S^{N}}} = \frac{1}{a^{2}\sin^{\ssc{N-1}}\phi_{1}\sin^{\ssc{N-2}}\phi_{2}...\sin\phi_{\ssc{N-1}}}\sum_{i=1}^{\ssc{N}}
\frac{\partial}{\partial \phi_{i}} \frac{\sin^{\ssc{N-1}}\phi_{1}\sin^{\ssc{N-2}}\phi_{2}...\sin\phi_{\ssc{N-1}}}{\left[\prod_{j=1}
^{i-1}\sin \phi_{j}\right]^{2}}\frac{\partial}{\partial \phi_{i}}\;.
\eeq  
The $S^{\ssc{N}}$ geometry is characterized by $N$ angles, $(N-1)$ of which
run from $0$ to $\pi$ ($\phi_{1},\ldots,\phi_{\ssc{N-1}}$). The azimuthal angle $\phi_{\ssc{N}}$ varies from $0$ to $2\pi$.
Introducing $\nabla^{2}_{\ssc{S^{N}}} = \frac{1}{a^{2}}\nabla^{2}$, Eq.~(\ref{spherewave}) becomes 
\beq
\label{equationN}
\left( \nabla^{2} + m_{\{n\}}^{2}a^{2} \right) \psi^{\{n\}}(\{\phi_{i}\}) = 0,
\eeq
which we solve using the ansatz
\beq
\psi^{\{n\}}(\{\phi_{i}\})=\psi_{1}(\phi_{1})\psi_{2}(\phi_{2})\ldots\psi_{\ssc{N}}(\phi_{\ssc{N}}).
\eeq
Eq.~(\ref{equationN}) can then be recast in the following form,
\begin{eqnarray}
\label{generalN}
\frac{1}{\psi_{1}}\frac{1}{\sin^{\ssc{N-1}}\phi_{1}}\frac{\partial}{\partial \phi_{1}} \sin^{\ssc{N-1}}\phi_{1} \frac{\partial}{\partial
\phi_{1}}\psi_{1} + m^{2}a^{2} + \frac{1}{\sin^{2}\phi_{1}} \left\{ \frac{1}{\psi_{2}}\frac{1}{\sin^{\ssc{N-2}}\phi_{2}}
\frac{\partial}{\partial \phi_{2}}\sin^{\ssc{N-2}}\phi_{2}\frac{\partial}{\partial \phi_{2}}\psi_{2} \right.\\
\left.+\frac{1}{\sin^{2}\phi_{2}} \left[ \frac{1}{\psi_{3}}\frac{1}{\sin^{\ssc{N-3}}\phi_{3}}
\frac{\partial}{\partial \phi_{3}}\sin^{\ssc{N-3}}\phi_{3}\frac{\partial}{\partial \phi_{3}}\psi_{3} + \ldots
+ \frac{1}{\sin^{2}\phi_{\ssc{N-1}}} \left( \frac{1}{\psi_{\ssc{N}}}\frac{\partial^{2}}{\partial \phi_{\ssc{N}}^{2}}
\psi_{\ssc{N}} \right) \ldots 
\right] \right\} \nonumber
 = 0.
\end{eqnarray}

Using the change of variables $x_{i}=\cos \phi_{i}$ and introducing the parameter $2\alpha_{i}=N-i$, Eq.~(\ref{generalN}) can be 
written as a set of $N$ differential equations:
\begin{eqnarray}
\label{gegenbauer}
(1-x_{1}^{2})\frac{\partial^{2}\psi_{1}}{\partial \phi_{1}^{2}} - (2\alpha_{1}+1)x_{1}\frac{\partial \psi_{1}}{\partial \phi_{1}}
+ \left[ n_{1}(n_{1}+2\alpha_{1}) - \frac{n_{2}(n_{2}-1+2\alpha_{1})}{1-x_{1}^{2}} \right] \psi_{1} &=& 0 \nonumber \\
&\vdots& \nonumber \\
(1-x_{i}^{2})\frac{\partial^{2}\psi_{i}}{\partial \phi_{i}^{2}} - (2\alpha_{i}+1)x_{i}\frac{\partial \psi_{i}}{\partial \phi_{i}}
+ \left[ n_{i}(n_{i}+2\alpha_{i}) - \frac{n_{k}(n_{k}-1+2\alpha_{k})}{1-x_{i}^{2}} \right] \psi_{i} &=& 0 \\ \nonumber
&\vdots& \nonumber \\
\label{gegenbauer2}
\frac{\partial^{2}\psi_{\ssc{N}}}{\partial \phi_{\ssc{N}}^{2}}+n_{\ssc{N}}^{2}\psi_{\ssc{N}}&=&0,
\end{eqnarray}
where the quantum numbers $n_{i}$ are such that $n_{1}\geq n_{2} \geq \cdots \geq n_{\ssc{N-1}}$ with $n_{i}=0,1,\ldots,\infty$ 
and the range of the azimuthal quantum number is $-n_{\ssc{N-1}}\leq n_{\ssc{N}}\leq n_{\ssc{N-1}}$. The mass spectrum is uniquely controlled by
$n_{1}$,
\beq
m^{2}=\frac{n_{1}(n_{1}+2\alpha_{1})}{a^{2}}=\frac{n_{1}(n_{1}+N-1)}{a^{2}}.
\eeq
The solutions to Eqs.~(\ref{gegenbauer}) can be expressed as a product of normalized
associated Gegenbauer polynomials \cite{gegen},
\beq
\psi^{(\alpha_{i})\;n_{k}}_{n_{i}}(x_{i}) = N^{(\alpha_{i})\;n_{k}}_{n_{i}}(1-x_{i}^{2})^{n_{k}/2}\frac{d^{n_{k}}}
{dx_{i}^{n_{k}}}C^{(\alpha_{i})}_{n_{i}}(x_{i}),
\eeq
where the $N^{(\alpha_{i})\;n_{k}}_{n_{i}}$'s are normalization constants and
\beq
C_{n}^{(\alpha)}(x)=\frac{\left(-1\right)^{n}}{2^{n}}\frac{\Gamma\left(2\alpha+n\right)
\Gamma\left(\frac{2\alpha+1}{2}\right)}{\Gamma\left(2\alpha\right)
\Gamma\left(\frac{2\alpha+2}{2}+n\right)}\frac{(1-x^{2})^{1/2-\alpha}}{n!}\frac{d^{n}}{dx^{n}}\left[ (1-x^{2})
^{\alpha+n-1/2}\right].
\eeq
The Gegenbauer polynomials can also be obtained from the hyper-spherical generating functional,
\beq
\frac{1}{(1-2xt+t^{2})^{\alpha}} = \sum_{n=0}^{\infty}C_{n}^{(\alpha)}(x)t^{n}.
\eeq
Using the measure on a $N$-sphere of radius $a$,
\beq
dV = a^{\ssc{N}}\sin^{\ssc{N-1}}\phi_{1}\sin^{\ssc{N-2}}\phi_{2}\ldots\sin\phi_{\ssc{N-1}},
\eeq 
the normalization constants are found to be
\beq
N^{(\alpha_{i})\;n_{k}}_{n_{i}} = \frac{\Gamma(\alpha_{i}+n_{k})}{2^{n_{k}}\alpha_{i}(\alpha_{i}+2)...(\alpha_{i}+n_{k
-1})}\left[ \frac{(n_{i}-n_{k})!(\alpha_{i}+n_{i})}{\pi \; 2^{(1-2\alpha_{i}-2n_{k})}\Gamma(2\alpha_{i}+n_{i}+n_{k})}
\right]^{1/2}. 
\eeq

For later reference we evaluate the graviton wave function at a given point on $S^{\ssc{N}}$.
For simplicity, we choose to evaluate it at ${\bf y}={\bf 0}$ which on a $N$-dimensional sphere
can be taken to correspond to $\phi_{1}=0$ with $\phi_{2} \ldots \phi_{\ssc{N}}$ being irrelevant variables. We use
the following property of the associated Gegenbauer polynomials at $x_{i}=\cos \phi_{i}=1$,
\beq
\psi^{(\alpha_{i})\;n_{k}}_{n_{i}}(1)=\frac{(2\alpha_{i}+n_{i}-1)!}{n_{i}!(2\alpha_{i}-1)!}\delta_{0}^{n_{k}},
\eeq
where $\delta_{0}^{n_{k}}$ is the Kronecker delta setting $n_{k}$ to zero. The wave function evaluated at a specific 
point can then be shown to be proportional to $\delta_{0}^{n_{2}}\delta_{0}^{n_{3}}\ldots
\delta_{0}^{n_{\ssc{N}}}$. Consequently, in a mode expansion of the graviton on a $N$-sphere only the $n_{1}$ quantum
number is seen to play a role. One might be led to think that this greatly reduces the density of states
allowed to propagate on the sphere but this
turns out not to be the case. In fact, the multiplicity at each quantum level $n_{1}$, which is $|\psi^{n_{1}}({\bf y}={\bf 0})|^{2}V_{\ssc{S^{N}}}$, 
reappears in the coupling
terms of SM matter with the graviton modes through the normalization constant. For later use, we write down the graviton wave function at
a specific point on the sphere
for $N=2$ to $6$ (from now on we use the convention $n_{1}=n$),
\beq
\label{pi}
\psi^{n}({\bf y}={\bf 0}) =  \left[ \frac{2n+1}{V_{S^{\ssc{2}}}} \right]^{1/2}, \;\;\;\; N=2
\eeq
\beq
\psi^{n}({\bf y}={\bf 0}) =  \left[ \frac{(n+1)^{2}}{V_{S^{\ssc{3}}}} \right]^{1/2}, \;\;\;\; N=3
\eeq
\beq
\psi^{n}({\bf y}={\bf 0}) =  \left[ \frac{(2n+3)(n+2)(n+1)}{6V_{S^{\ssc{4}}}}\right]^{1/2}, \;\;\;\; N=4
\eeq
\beq
\psi^{n}({\bf y}={\bf 0}) =  \left[ \frac{(n+3)(n+2)^{2}(n+1)}{12V_{S^{\ssc{5}}}}\right]^{1/2}, \;\;\;\; N=5
\eeq
\beq
\label{pf}
\psi^{n}({\bf y}={\bf 0}) =  \left[ \frac{(5+2n)(n+4)(n+3)(n+2)(n+1)}{120V_{S^{\ssc{6}}}}\right]^{1/2}, \;\;\;\; N=6
\eeq 
where $V_{S^{\ssc{N}}}=\frac{2\;\pi^{(N+1)/2}a^{\ssc{N}}}{\Gamma\left((N+1)/2\right)}$ is the volume of a $N$-sphere. As will
be made clear in Sec.~\ref{emission}, the wave function evaluated at a given point (directly related to the multiplicity 
at each KK level) 
is crucial for
understanding how processes on the three-brane are affected by the geometry of the internal space.
   
\section{Probing the Extra Dimensions}
\label{emission}

We now consider in detail how the geometry of the internal space affects the couplings of SM fields to the gravitational sector.
The linear coupling, which is universally determined by general covariance,
is of the form (see Ref.~\cite{lykken})
\beq
\label{couplingbrane}
\frac{1}{M_{\ssc{D}}^{\ssc{1+\frac{N}{2}}}}\int d^{\ssc{D}}x \; h^{\ssc{AB}}T_{\ssc{AB}},
\eeq
where $T_{\ssc{AB}}$ is the stress-energy tensor associated with SM fields on the three-brane. 
We are studying gravity on the product space $M^{\ssc{4}}\times B^{\ssc{N}}$ with SM fields localized on the 
$M^{\ssc{4}}$ submanifold. 
It is a reasonable approximation for our purposes to use the following form for the stress-energy tensor:
\beq
\label{st}
T_{\ssc{AB}}(x,{\bf y})=\delta_{\ssc{A}}^{\mu}\,\delta_{\ssc{B}}^{\nu}T_{\mu\nu}(x)\delta^{\ssc{(N)}}({\bf y}).
\eeq
This expression is written in the so-called static gauge which consists in ascribing four bulk coordinates to the three-brane
($A=0,1,2,3 \rightarrow \mu=0,1,2,3$) and the remaining $N$ coordinates to the internal space ($A=4,...,D-1 \rightarrow i=1,...,N$). 
Using Eq.~(\ref{st}) is equivalent to considering an infinitely thin and tensionless brane. A consequence
of this simplification is that tree level diagrams involving the exchange of off-shell gravitons are not finite\footnote{When
constructing the effective interactions due to virtual KK modes exchange, an infinite sum needs to be evaluated. It
is in most cases a divergent quantity unless one introduces by hand an ultraviolet cut-off. 
This is not a very natural way to cure the problem as the cut-off remains in the final expression.} which
goes against intuition. In fact, we expect loop diagrams to diverge but not the tree level ones. Let us pause
and consider this problem. The incoming and outgoing states in a typical process are SM fermions that are confined 
to the flat submanifold $M^{\ssc{4}}$. From the point of view of an observer on the brane, a graviton
is emitted at one vertex, propagates into the bulk, and is reabsorbed at the second vertex. Using a stress-energy tensor 
of the form (\ref{st}) to work out the expression for the tree level amplitude shows that momentum is conserved on the
brane but not in the internal space. In other words, there is no constraint at the vertices on the transverse 
momentum of the graviton. This is similar to what happens in a loop diagram so one should not be surprised 
that tree level amplitudes may diverge. This puzzle is resolved in Ref.~\cite{bando} where the authors give the 
brane a finite 
tension and take into account its fluctuation modes in the transverse directions. This induces an exponential factor at the
vertices which naturally cuts off the problematic ultraviolet modes responsible for the divergences. 
This phenomenon is of 
no concern to us since we are only considering finite interactions involving on-shell gravitons. Nevertheless, such a
suppression factor for KK modes emitted from the three-brane should be included in a thorough analysis. 

Using Eqs.~(\ref{couplingbrane}) and (\ref{st}) insures that everything coupling to the gravitational sector is located on 
the three-brane. Using the ansatz (\ref{simpleansatz}), the coupling term becomes
\begin{eqnarray}
S^{\ssc{L}}_{\ssc{M}} = \frac{1}{M_{\ssc{D}}^{\ssc{1+\frac{N}{2}}}}\int d^{\ssc{N}}y \;\delta^{\ssc{(N)}}({\bf y}) \int d^{4}x \;T_{\mu\nu}(x)
\delta_{\ssc{A}}^{\mu}\delta_{\ssc{B}}^{\nu}
h^{\ssc{AB}}(x,{\bf y}) \nonumber
\end{eqnarray}
\begin{eqnarray}\scriptsize
= \frac{1}{M_{\ssc{D}}^{\ssc{1+\frac{N}{2}}}} \int d^{4}x \; Tr \left(
\begin{array}{cc}
\sum h_{\mu\nu}^{\{n\}}(x)\psi^{\{n\}}({\bf y}={\bf 0}) & 0 \\
0 & 0
\end{array}
\right) \left(
\begin{array}{cc}
T^{\mu\nu}(x) & 0 \\
0 & 0
\nonumber
\end{array}
\right).
\end{eqnarray}
Consequently, each KK mode is characterized by the coupling
\beq
\label{coupling}
\frac{1}{M_{\ssc{D}}^{\ssc{1+\frac{N}{2}}}}\psi^{\{n\}}({\bf y}={\bf 0})\int d^{4}x \; h_{\mu\nu}^{\{n\}}(x)T^{\mu\nu}(x).
\eeq
When the compact geometry is a torus this expression becomes
\beq
\label{toruscoupling}
\frac{1}{M_{\ssc{P}}}\int d^{4}x \; h_{\mu\nu}^{{\bf n}}(x)T^{\mu\nu}(x),
\eeq
where use has been made of Eq.~(\ref{toricw}) and the fact that $M_{\ssc{P}}=V_{\ssc{N}}^{\ssc{1/2}}M_{\ssc{D}}^{\ssc{1+N/2}}$. 
When considering a spherical compact geometry, we obtain the
same kind of expression: 
\beq
\label{spherecoupling}
\frac{f_{\ssc{N}}(n)}{M_{\ssc{P}}}\int d^{4}x \; h_{\mu\nu}^{n}(x)T^{\mu\nu}(x),
\eeq  
where $f_{\ssc{N}}(n)$ represents a family of polynomials in $n$ (see Eqs.~(\ref{pi})-(\ref{pf})) related to
the mutliplicity of the states propagating on the sphere at each KK level.  

Using Eq.~(\ref{coupling}) we can find the Feynman rules 
for processes involving gravitons coupled to SM fields on the three-brane \cite{lykken,wells}. We restrict 
ourselves to studying the potential 
relevance of the process $e^{+}e^{-}\rightarrow \gamma h$ for probing the geometry of the transverse space. From the
four-dimensional point of view on submanifold $M^{\ssc{4}}$ this corresponds to the emission of a kinematically cut-off tower
 of massive graviton modes
during a high energy collision. 
The differential cross section for the emission of a
{\rm QED} photon and a massive graviton (denoted $h_{m}$) following an $e^{+}e^{-}$ collision with CM energy $\sqrt{s}$ is
\cite{wells,peskin}
\beq
\label{diffgrav}
\frac{d\sigma_{m}}{dt}(e^{+}e^{-}\rightarrow \gamma h_{m})=\frac{\alpha}{16}\frac{1}{M_{\ssc{P}}^{2}}F(x,y),
\eeq
where $\alpha$ is the electromagnetic fine-structure constant and
\beq
\label{diffgrav2}
F(x,y)=\frac{-4x(1+x)(1+2x+2x^{2})+y(1+6x+18x^{2}+16x^{3})-6y^{2}x(1+2x)+y^{3}(1+4x)}{x(y-1-x)}
\eeq
with $x=t/s$ and $y=m^{2}/s$. From our limited four-dimensional point of view we do not distinguish
between gravitons of different transverse momenta (mass-squared). Thus, the actual cross section for graviton emission
from the brane is obtained by summing Eq.~(\ref{diffgrav}) over all kinematically allowed values of $m^{2}$ $\ie$ 
up to $m^{2}=s$. Note that when we use the variable $y=m^{2}/s$ the sum conveniently runs from $y=0$ to $1$.

From an experimental perspective, the $e^{+}e^{-}\rightarrow \gamma h$ process is competing with the Standard Model 
background $e^{+}e^{-}\rightarrow \gamma \nu \bar{\nu}$. Of course, when there are either no or extremely 
small extra dimensions the gravitational
process, being suppresed by a $M_{\ssc{P}}^{-2}$ factor, is completely undetectable. With large extra dimensions 
the relatively important
number of KK modes enhances the graviton signal which leads to a potentially detectable departure from the
SM signature. This corresponds to
$\sigma (e^{+}e^{-}\rightarrow \gamma h)$ no longer being suppressed by a $M_{\ssc{P}}^{-2}$ factor but by a 
$M_{\ssc{D}}^{\ssc{-(2+N)}}$ factor. Consequently, picking $M_{\ssc{D}}$ as small as possible leads to larger 
gravitational signals. There
is a fundamental limitation in our freedom to do that though. In fact, using the Gauss law one finds the following low
energy constraint\cite{ADD}:
\beq
\label{constraint}
M_{\ssc{P}}^{2} = M^{\ssc{2+N}}_{\ssc{D}}V_{\ssc{N}},
\eeq   
where $V_{\ssc{N}}$ is the volume of the compact space. Requiring the effective low energy four-dimensional gravitational coupling
to be the observed Newton constant $G_{\ssc{N}}$ is equivalent to imposing Eq.~(\ref{constraint}) which is a relationship between
the size of the extra dimensions, their number, $N$, and the true quantum gravity scale, $M_{\ssc{D}}$. It is interesting to 
note that experiments have been performed
probing gravity down to approximately one millimeter \cite{adel} without finding any discrepancies with the usual $1/r$ 
potential. 
Based on Eq.~(\ref{constraint}), this implies that for $N=2$ the
quantum gravity scale could be as low a 1 {\rm TeV}. Although this particular set of parameters seems to be excluded 
by astrophysical constraints \cite{astro}, it does not mean that other values of $N$ and $M_{\ssc{D}}$ leading to 
detectable signatures have to be rejected.
It should be clear that the graviton signal is increasingly suppressed relative to the 
background process $e^{+}e^{-}\rightarrow \gamma \nu \bar{\nu}$ as $M_{\ssc{D}}$ and $N$ are increased.

\subsection{Phase Space Integrals on $T^{\ssc{N}}$}
For a $T^{\ssc{N}}$ geometry the wave function for the transverse graviton modes at ${\bf y}={\bf 0}$ is independent of ${\bf n}$. Based on
Eq.~(\ref{toruscoupling}) this means that there is no restriction on the quantum numbers of the modes that are emitted at 
a given point on the torus (from the three-brane). 
Then the operator we use to sum over transverse momenta is \cite{peskin}
\beq
\label{sumtorus}
{\cal O}_{\ssc{T^{N}}} = \sum_{\ssc{{\bf k}}} \rightarrow \frac{1}{V_{\ssc{|{\bf k}|}}}\int d^{\ssc{N}}k = \frac{\Omega_{\ssc{N}}
R^{\ssc{N}}}{2} \int m^{\ssc{N-2}} dm^{2} = \frac{\Omega_{\ssc{N}}R^{\ssc{N}}s^{\ssc{N/2}}}{2}\int_{0}^{1}dy\;y^{\ssc{(N-2)/2}}, 
\eeq
where $V_{\ssc{|{\bf k}|}}$ is the volume occupied by one state in ${\bf k}$-space, $y=m^{2}/s$ and
\beq
\Omega_{\ssc{N}}=\frac{2\;\pi^{\ssc{N/2}}}{\Gamma(N/2)}
\eeq
is the volume of a unit sphere embedded in a $N$-dimensional space. Because the extra 
dimensions are assumed to be large (in {\rm TeV}$^{-1}$ units) with respect to the inverse center of mass energy of the 
process, it is reasonable to take the continuum limit when
performing the sum 
over momenta (see the Appendix). For future comparison with $S^{\ssc{N}}$ phase space integrals we explicitly write down Eq.~(\ref{sumtorus}) for
$N=2$ to $6$:
\beq
\label{ti}
{\cal O}_{T^{2}}\rightarrow V_{T^{2}}\frac{s}{4\pi}\int_{0}^{1}dy
\eeq
\beq 
{\cal O}_{T^{3}}\rightarrow V_{T^{3}}\frac{s^{3/2}}{4\pi^{2}}\int_{0}^{1}dy \; y^{1/2}
\eeq
\beq
{\cal O}_{T^{4}}\rightarrow V_{T^{4}}\frac{s^{2}}{16\pi^{2}}\int_{0}^{1}dy \; y
\eeq
\beq
{\cal O}_{T^{5}}\rightarrow V_{T^{5}}\frac{s^{5/2}}{24\pi^{3}}\int_{0}^{1}dy \; y^{3/2} 
\eeq
\beq
\label{tf}
{\cal O}_{T^{6}}\rightarrow V_{T^{6}}\frac{s^{3}}{128\pi^{3}}\int_{0}^{1}dy \; y^{2}.
\eeq
To evaluate the differential cross section for the emission of a $(4+N)$-dimensional graviton
from the three-brane we need only apply
operator ${\cal O}_{\ssc{T^{N}}}$ to Eq.~(\ref{diffgrav}). As an example 
Fig.~1 shows the total cross sections (after the integration over angles has been performed) 
as a function of $s$ for graviton 
emission when $M_{\ssc{D}}=1$ {\rm TeV} for $N=2$ to $6$.  

\begin{figure}[t]
\label{fig1}
\begin{center}

\epsfig{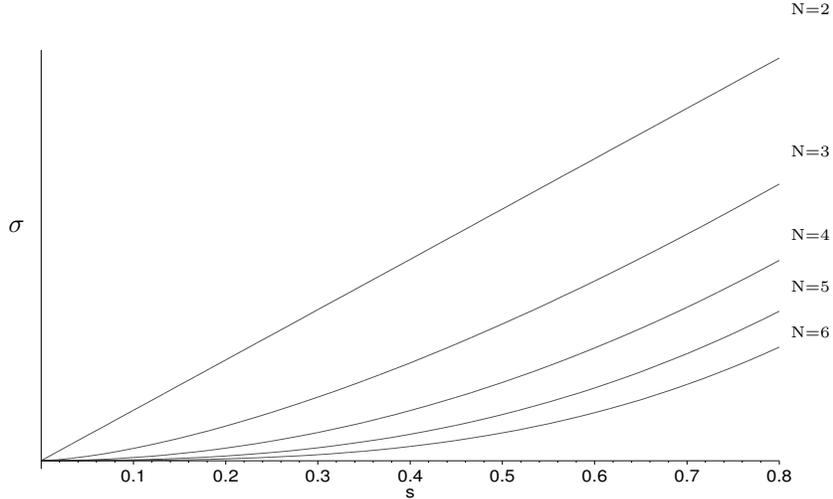}

\end{center}

\begin{picture}(0,0)
\put(31,45){{\footnotesize $\sigma$}}
\put(135,74){\tiny N=2}
\put(135,55){\tiny N=3}
\put(135,44){\tiny N=4}
\put(135,37){\tiny N=5}
\put(135,31){\tiny N=6}
\end{picture}

\caption{Cross sections (arbitrary units) for $e^{+}e^{-}\rightarrow \gamma h$, in models with the extra dimensions compactified 
on $T^{\ssc{N}}$, as a function of $s$ (in {\rm TeV}$^{2}$) for $M_{\ssc{D}}=1$ {\rm TeV}. Starting 
from the top, curves are for $N=2$ to 6.}
\end{figure}

\subsection{Phase Space Integrals on $S^{N}$}
As shown in Sec.~\ref{psi} the graviton wave function evaluated at a given point on $S^{\ssc{N}}$ only depends on
the quantum number $n$. Using Eqs.~(\ref{pi})-(\ref{pf}), it is straightforward to write down the operators analogous
to Eqs.~(\ref{ti})-(\ref{tf}) when the compact geometry is $S^{\ssc{N}}$,
\beq
\label{si}
{\cal O}_{S^{2}}\rightarrow V_{S^{2}}\frac{s}{4\pi}\int_{0}^{1}dy
\eeq
\beq
{\cal O}_{S^{3}}\rightarrow V_{S^{3}}\frac{s^{3/2}}{4\pi^{2}}\int_{0}^{1}dy \left(y+\frac{1}{a^{2}s}\right)^{1/2}
\eeq
\beq
{\cal O}_{S^{4}}\rightarrow V_{S^{4}}\frac{s^{2}}{16\pi^{2}}\int_{0}^{1}dy \left(y+\frac{2}{a^{2}s}\right)
\eeq
\beq
{\cal O}_{S^{5}}\rightarrow V_{S^{5}}\frac{s^{5/2}}{24\pi^{3}}\int_{0}^{1}dy \left(y+\frac{3}{a^{2}s}\right)
\left(y+\frac{4}{a^{2}s}\right)^{1/2} 
\eeq
\beq
\label{sf}
{\cal O}_{S^{6}}\rightarrow V_{S^{6}}\frac{s^{3}}{128\pi^{3}}\int_{0}^{1}dy \left(y+\frac{4}{a^{2}s}\right)
\left(y+\frac{6}{a^{2}s}\right).
\eeq
The parameter playing a role in distinguishing a spherical from a toric geometry depends on the size of the internal 
space. We label it
\beq
\label{max}
d_{\ssc{a}}=\frac{1}{n_{\ssc{max}}^{2}},
\eeq 
where $n_{\ssc{max}}=a\sqrt{s}$ is (if $N \ll n_{\ssc{max}}$) the maximum quantum number 
over which we integrate when performing the phase space sum on a sphere.
If we integrate over an overwhelmingly large number of states ($d_{\ssc{a}}\rightarrow 0$) the 
$e^{+}e^{-}\rightarrow \gamma h$ cross section 
evaluated on $T^{\ssc{N}}$ is expected to be very close to the one evaluated on $S^{\ssc{N}}$. 
This corresponds to a sector of the theory for which 
the typical size of the extra dimensions (in {\rm TeV}$^{-1}$ units) is large with respect to the inverse center
of mass energy of the process ($a \gg 1/\sqrt{s}$). In this case the spacing between KK levels is small compared with
the CM energy.
For example, it can be seen that $a\sim 2\times10^{5}$ {\rm TeV}$^{-1}$ if $N=6$, $M_{\ssc{D}}=1$ {\rm TeV} and $a\sim 
200$ {\rm TeV}$^{-1}$ with $M_{\ssc{D}}=30$ {\rm TeV}. As we increase both $M_{\ssc{D}}$ and $N$ we expect 
the difference between
cross sections on $T^{\ssc{N}}$ and $S^{\ssc{N}}$ to increase since this corresponds to taking larger values of $d_{\ssc{a}}$
(a smaller number of KK modes are summed over).
The numerical factors multiplying $d_{\ssc{a}}$ in ${\cal O}_{\ssc{S^{N}}}$ are more important for large $N$
which also contributes in enhancing the difference in cross sections between the two geometries.   

When
the internal space has a typical length scale which is extremely small (with respect to the inverse CM energy of the process),
one expects processes taking place on a torus
to be indistinguishable from processes on a sphere (or on any other smooth manifold for that matter). This is not reflected 
in our phase space integral procedure. In the limit when the extra dimensions are extremely small
the procedure we are using is not valid anymore since then it is highly probable that, for
the range of CM energies considered, only the zero-mode of the graviton will be excited. The phase space integral procedure is 
useful
only when numerous modes are excited $\ie$ when the extra dimensions are large.   
The $N=2$ case is special since 
it is then impossible, for large extra dimensions, to distinghish between the $T^{2}$ and the $S^{2}$ geometries
using a process 
such as $e^{+}e^{-}\rightarrow \gamma h$ (${\cal O}_{\ssc{T^{2}}}={\cal O}_{\ssc{S^{2}}}$).

If $d_{\ssc{a}}$ is not too small we expect differences in cross sections evaluated on $T^{\ssc{N}}$ and
$S^{\ssc{N}}$ for $N>2$. Comparing Eqs.~(\ref{ti})-(\ref{tf}) and Eqs.~(\ref{si})-(\ref{sf}) shows that models 
with a spherical transverse space 
will lead to larger cross sections. 
Fig.~2 represents the behavior of cross sections for the two geometries studied when $M_{\ssc{D}}=30$ {\rm TeV} and $N=6$. 
We see that as
$s$ is augmented the difference between the cross sections slightly increases. This suggests that overall the
number of KK modes excited on a $N>2$ sphere is larger than on a torus. Based on Eq.~(\ref{max}) we see that $s$ is related 
to the maximum quantum number ($n_{\ssc{max}}$) over which the integration is performed. Consequently, the larger the CM energy is, the larger we expect the 
cross section differences to be (a large $s$ corresponds to integrating over more modes). Since highly energetic modes are not expected to differentiate between smooth geometries
(their wavelength is assumed to be much smaller than the inverse curvature-squared of the internal space), there exists a CM energy beyond which
the multiplicity at each level is the same both for the sphere and the torus\footnote{We do not specify what this critical value
of $s$ is as it depends both on $M_{\ssc{D}}$ and $N$. It represents a natural separation between the low and 
high energy modes propagating on the compact manifold. For example, when $M_{\ssc{D}}=30$ {\rm TeV} and $N=6$, the critical
value is around $s=0.15$ {\rm TeV}$^{2}$.}. Past this critical 
$s$-value, we expect the difference between cross sections to become constant. Although this is not obvious from Fig.~3, we have shown numerically that this is in fact what happens.  

\begin{figure}[t]
\label{sec1}
\begin{center}
\epsfig{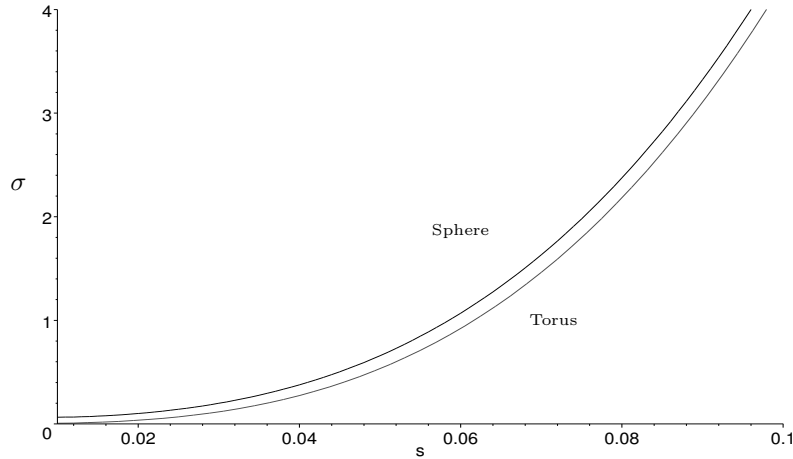}
\end{center}

\begin{picture}(0,0)
\put(31,45){{\footnotesize $\sigma$}}
\put(87,39){\tiny Sphere}
\put(100,27){\tiny Torus}
\end{picture}

\caption{Comparison of cross sections (in units of $[\alpha/\pi^3 M_{\ssc{D}}^{8}]$) for $e^{+}e^{-}\rightarrow \gamma h$ 
in models with compact manifold $T^{6}$ and $S^{6}$ 
when $M_{\ssc{D}}=30$ {\rm TeV} for $s$ varying from 0.01 to 0.1 {\rm TeV}$^{2}$. The upper curve corresponds to an internal
space with a spherical geometry and the lower
curve to a toric geometric. The difference between the two curves slightly increases with $s$.}
\end{figure}

Approximating sums over graviton modes by integrals is valid for the range
of $M_{\ssc{D}}$ we are studying because the spacing between quantum levels (in momentum space) is small. 
In fact, this is of the order of magnitude $1/a$ where, for example, $a \approx 2\times 10^{5}$ {\rm TeV}$^{-1}$ when $N=6$ 
and $M_{\ssc{D}}
=1$ {\rm TeV}. As we increase $M_{\ssc{D}}$ the typical size of the extra dimensions decreases therefore leading to a larger
spacing. This has the potential of invalidating the sums by integrals approximation for sufficiently large values of the
quantum gravity scale. For more details on potential errors induced by the approximation see the Appendix.    

\begin{figure}[t]
\label{sec2}
\begin{center}
\epsfig{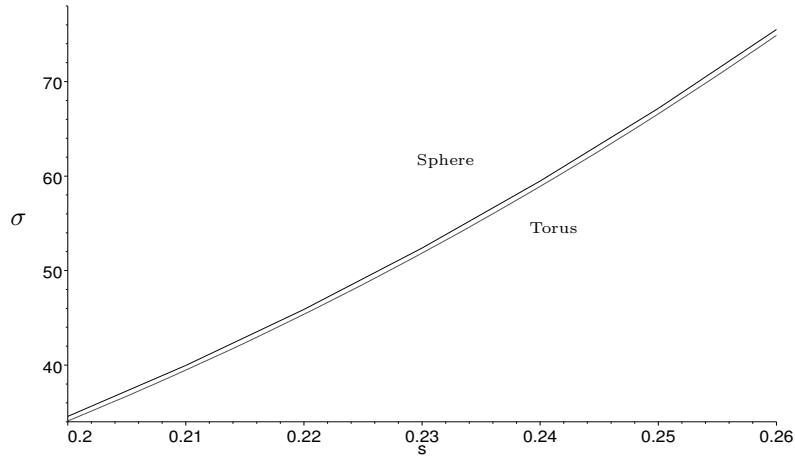}
\end{center}

\newcounter{xc2}
\newcounter{yc2}
\begin{picture}(0,0)
\put(31,40){{\footnotesize $\sigma$}}
\put(85,48){\tiny Sphere}
\put(100,39){\tiny Torus}
\end{picture}

\caption{Comparison of cross sections (in units of $[\alpha/\pi^3 M_{\ssc{D}}^{8}]$) for $e^{+}e^{-}\rightarrow \gamma h$ 
in models with compact manifold $T^{6}$ and $S^{6}$ 
when $M_{\ssc{D}}=30$ {\rm TeV} for $s$ varying from 0.2 to 0.3 {\rm TeV}$^{2}$. The upper curve corresponds to an internal
space with a spherical geometry and the lower
curve to a toric geometric. Beyond a critical value of $s$ (around $s=0.15$ {\rm TeV}$^{2}$ in this case) the difference between 
the two curves becomes constant.}
\end{figure}

A parameter we can use to quantify the effect of the compact geometry on graviton emission from the three-brane is the ratio
\beq
D_{\ssc{N}}(s,M_{\ssc{D}})= \frac{\sigma_{\ssc{S^{N}}}(e^{+}e^{-}\rightarrow \gamma h)-\sigma_{\ssc{T^{N}}}(e^{+}e^{-}\rightarrow \gamma h)}
{\sigma_{\ssc{S^{N}}}(e^{+}e^{-}\rightarrow \gamma h)},
\eeq     
which is a function of the size of the extra dimensions through $N$ and $M_{\ssc{D}}$. This quantity characterizes relative 
rather than absolute differences. As shown earlier, 
the ratio $D_{\ssc{N}}$
is zero for $N=2$.
For $N=3$ and $N=4$, all considered values of $M_{\ssc{D}}$ and $\sqrt{s}$
lead to a function $D_{\ssc{N}}$ which is a negligible fraction of a percent. $D_{5}$ is also negligible for $M_{\ssc{D}}
=1$ {\rm TeV} but reaches $0.002\%$ for $M_{\ssc{D}}=10$ {\rm TeV} (for small values of $s$). When $M_{\ssc{D}}=30$ {\rm TeV} 
the ratio $D_{5}$ goes as high 
as $0.04\%$ 
for $\sqrt{s}=0.1$ {\rm TeV} but 
goes down to $0.002\%$ for $\sqrt{s}=0.5$ {\rm TeV}. The most noticeable effects occur for $N=6$. Then,
with $M_{\ssc{D}}=30$ {\rm TeV}, $D_{6}$ varies from $5\%$ to $0.1\%$ as $\sqrt{s}$ spans the $0.1$ to $0.5$ {\rm TeV} 
range. Still for $N=6$, when 
$M_{\ssc{D}}=10$ {\rm TeV}
$D_{\ssc{N}}$ varies from $0.3\%$ to $0.01\%$ but is negligible for $M_{\ssc{D}}=1$ {\rm TeV}. 

In conclusion, we find that the relative difference between cross sections (for a given $s$) in models with spherical 
and toric geometries takes larger values when the quantum gravity scale is large and the dimensions are numerous. While
the absolute difference increases with $s$ (for a given $M_{\ssc{D}}$ and $N$) the relative difference, $D_{\ssc{N}}$, does
just the opposite. This is expected as cross sections are rising functions of the CM energy.  
   
\section{Concluding Remarks} 
\label{dis}

The contribution of each KK mode to the effective four-dimensional gravitational potential
is proportional to 
\beq
\frac{e^{-(const.)r}}{r},
\eeq
where the constant is related to the mass of the mode. In the large $r$ limit the potential
takes the following form:
\beq
V(r) \simeq -\frac{1}{r}(1 + \kappa e^{-m_{\{1\}}r}),
\eeq 
where $\kappa$ is the multiplicity of the first massive KK mode and $m_{\{1\}}$ its mass. It can be shown 
that $\kappa$ is somewhat larger for a torus (with all compactification radii assumed to be the same) than it is for a sphere. 
This suggests that to an observer on the three-brane the force
due to gravity is slightly stronger if the extra dimensions are toric. This is not in contradiction with our result 
that cross sections for graviton emission associated with a spherical manifold are larger than those evaluated with a toric 
internal space. While it is true that low energy gravity is stronger for the torus,
it simply appears otherwise from a microscopic point of view when gravity is probed with high energies. It is argued in 
Ref.~\cite{devi} that large extra
dimensions compactified on some Calabi-Yau manifolds are such that $\kappa$ can be as large as 20 which is noticeably
larger than the corresponding multiplicities on the 6-sphere and the 6-torus. It is hopeless to try and find a generic solution
for multiplicities at all KK levels for Calabi-Yau manifolds as there exists an overwhelmingly large number of such spaces
(vacuum degeneracy problem of string theory). 
Nevertheless,
it is conceivable that models with large extra dimensions compactified on a Calabi-Yau might lead to more significant
discrepancies for microscopic processes compared with models where the internal space is either
toric or spherical.      
 
Having graviton modes propagating on a $N$-torus is exactly the same as having them existing in a 
$N$-dimensional box. Such a geometry has no intrinsic curvature so whether the modes have low or
high discretized momenta does not matter. By that we mean that all modes perceive the space as being $T^{\ssc{N}}$. If 
the compact geometry is $S^{\ssc{N}}$ the situation is different. The high energy modes, having a small
wavelength in transverse space, do not behave differently than when they are propagating on a $N$-torus (the wavelength-squared
is assumed to be much smaller that the inverse local curvature).
It then makes sense to say that the physics resulting from these high energy modes cannot be used to distinguish between processes 
taking place on different compact
geometries (unless the associated curvature is large). 
The graviton modes to which are associated small quantum numbers (low energy modes) are the ones that can 
be used to study the shape of the extra dimensions. In fact, their wavelength is presumably large enough to
allow them to recognize a sphere from a torus say. The multiplicity of states
(or density of states) at each quantum level on the compact geometry lattice grows as the norm of the momentum is increased.
We have shown that overall this mulitiplicity is larger on the sphere. This explains why the cross section for a process like $e^{+}e^{-}\rightarrow \gamma h$ is larger
when the compact geometry has a spherical symmetry. We have seen that past a certain 
transerve momenta the multiplicity on a sphere and a torus become equal. This means that beyond some critical value
for the CM energy, the difference between the cross sections evaluated for different geometries stabilizes to a constant
value. This is what we have found for the $N=6$, $M_{\ssc{D}}=30$ {\rm TeV} case but this is true in general. As $M_{\ssc{D}}$ is augmented the geometry of the compact
space plays an increasingly important role. While this is true, it is also worth noting that when $M_{\ssc{D}}$ is increased, 
deviations from the $e^{+}e^{-}\rightarrow \gamma \nu \bar{\nu}$ process progressively become negligible. In fact, the size of the
extra dimensions then becomes small which allows only a limited number of modes to propagate in the extra dimensions. 

In summary, high energy colliders are limited in their ability to determine the geometry of large extra dimensions.
This is due to two factors: 
\begin{itemize}
\item as $\sqrt{s}$ is increased the relative difference between
cross sections on different geometries decreases because the contribution of the low energy modes becomes increasingly small;
\item the impact of the geometry is more important for a large quantum gravity scale and numerous extra dimensions but this
also corresponds to a smaller gravitational contribution to SM background processes.
\end{itemize}   
In fact, the geometrical effect we find for the $e^{+}e^{-}\rightarrow \gamma h$ process is rather small. Nevertheless, it
is still conceivable that it could be detected at the upcoming {\rm LHC}, using
the process $q\bar{q}\rightarrow \gamma h$, for certain values of $N$ and $M_{\ssc{D}}$. This experimental study and the effects of
the geometry on supernova constraints will be considered in upcoming work \cite{leblond3}. \\
\\
\\
{\bf Acknowledgements:} The author is indebted to R.~C.~Myers for many insightful comments during the realization 
of this work and also wishes to thank C.~P.~Burgess, 
J.~M.~Cline and H.~Firouzjahi for useful discussions. This work was supported by the Natural Sciences and Engineering 
Research Council of Canada (NSERC). 

\section*{Appendix: Estimating Sums by Integrals}
We now verify that approximating sums over KK modes propagating into
the extra dimensions by integrals is valid for the range of parameters ($s$, $N$ and $M_{\ssc{D}}$) considered in this
paper. In order to do that we use the Euler-MacLaurin formula (see for example \cite{boas}) which we write
down schematically:
\begin{eqnarray}
\label{ugly}
\sum_{k} f(k) = \frac{1}{w}\int_{k_{min}}^{k_{max}}dk\;f(k) + \frac{1}{2} \left[ f(k_{min}) + f(k_{max}) \right] \\ \nonumber +
\sum_{s=\ssc{1}}^{m}\; \left[ \frac{B_{\ssc{2}s}}{(2s)!} w^{\ssc{2}s-1}
\left( f^{(\ssc{2}s-1)}(k_{min})-f^{(\ssc{2}s-1)}(k_{max})\right) + R_{s} \right],
\end{eqnarray} 
where $w$ characterizes the spacing between levels in the
Kaluza-Klein tower, the $B$'s are Bernouilli numbers ($B_{2}=1/6$, $B_{4}=-1/30$, $B_{6}=1/42$, $B_{8}=-1/30$, etc.)
and
\beq
R_{s} = w^{\ssc{2}s-1} \int_{k}d k\; f^{(\ssc{2}s+1)}(k) P_{\ssc{2}s+1}(k)
\eeq 
where $P_{\ssc{2}s+1}$ is a polynomial made of an infinite sum over oscillating functions of $k$, $k_{min}$ and $w$, the details
of which are not important for our purpose.

The mass of the KK modes propagating in the extra dimensions depends on integer-valued quantum numbers.
In the case of a space compactified on a 6-sphere the following sum needs to be evaluated,
\beq
\label{sumi}
\sigma_{\ssc{S^{6}}}(e^{+}e^{-}\rightarrow \gamma h) = \frac{\alpha}{16}\frac{1}{M_{\ssc{P}}^{2}}\sum_{n} F\left(x,m^{2}=
\frac{n(n+5)}{a^{2}}\right). 
\eeq
The function $F(x,y)$ where $y=m^{2}/s$ is introduced in Sec.~\ref{emission}. Using Eq.~(\ref{ugly}) it can be shown that the sum
in Eq.~(\ref{sumi}) 
is equivalent to an integral plus corrections parametrized by $\Delta_{\ssc{S^{6}}}$.
To a good approximation we can write
\beq
\label{correction1}
\Delta_{\ssc{S^{6}}} = \frac{1}{2\,s\,a^{2}} \left[ \frac{24}{s^{2}R^{4}} F(x,0) + \left( 1 + \frac{4}{s a^{2}} \right)
\left( 1+\frac{6}{s a^{2}} \right) F(x,1) \right]
\eeq
where there remains to be performed an angular integration ($x$-variable). The other corrections
include all order derivatives of $F(x,y)$ with respect to $y$ and are proportional to rising powers of $1/(\sqrt{s} a)$ the most
significant contribution being ${\cal O}(1 / s^{3/2}a^{3})$. Since $F(x,y)$ is a smooth slowly varying function of $y$ and
because $1 / s^{3/2}a^{3}$ is small (varies from 0.002 to $2\times 10^{-7}$ for $\sqrt{s}=0.1$ {\rm TeV}
to $\sqrt{s}=1$ {\rm TeV} when $M_{\ssc{D}}=30 \;{\rm TeV}$ and $N=6$) these contributions
are small compared with the one obtained using Eq.~(\ref{correction1}). The corresponding expression for a 6-torus is  
\beq
\label{correction2}
\Delta_{\ssc{T^{6}}} = \frac{1}{2\,s\,a^{2}} F(x,1).
\eeq
We note that the corrections (\ref{correction1}) and (\ref{correction2}) are more significant when $s$, $M_{\ssc{D}}$
and $N$ are large. We computed those explicitely for $N=6$, $M_{\ssc{D}}=30$ {\rm TeV} and $\sqrt{s}=0.1$ to 1 
{\rm TeV}. The absolute value of the corrections are to a good approximation equal for the spherical and toroidal
compactifications. Consequently, the corrections slightly rise up both curves in Fig.~2 and Fig.~3 by the
same amount for each given $s$. The relative importance of the corrections is aprroximately $1/5$ the value of the
relative differences we found between cross sections evaluated on a torus and on a sphere. Of course, as $M_{\ssc{D}}$
is increased the approximation becomes less precise as an inceasingly small number of modes are summed over.


\begin{thebibliography}{99}

\bibitem{ADD} N.~Arkani-Hamed, S.~Dimopoulos and G.~Dvali,
Phys.\ Lett.\ B {\bf 429}, 263 (1998) [hep-ph/9803315]; I.~Antoniadis, N.~Arkani-Hamed, S.~Dimopoulos and G.~Dvali,
Phys.\ Lett.\ B {\bf 436}, 257 (1998) [hep-ph/9804398].

\bibitem{astro} N.~Arkani-Hamed, S.~Dimopoulos and G.~Dvali, Phys.\ Rev.\ D {\bf 59}, 086004 (1999) [hep-ph/9807344];
S.~Nussinov and R.~Shrock, Phys.\ Rev.\ D {\bf 59}, 105002 (1998) [hep-ph/9811323]; 
S.~Cullen and M.~Perelstein,
Phys.\ Rev.\ Lett.\ {\bf 83}, 268 (1999) [hep-ph/9903422]; V.~Barger, T.~Han, C.~Kao and R.~J.~Zhang, Phys.\ Lett.\ B {\bf 461}, 34 (1999) [hep-ph/9905474];
D.~Atwood, C.~P.~Burgess, E.~Filotas, F.~Leblond, D.~London and I.~Maksymyk, Phys.\ Rev.\ D {\bf 63}, 025007 
(2001) [hep-ph/0007178]; C.~Hanhart, D.~R.~Phillips, S.~Reddy and M.~J.~Savage, Nucl.\ Phys.\ {\bf B595}, 335 (2001) [nucl-th/0007016]; M.~Fairbairn,
"Cosmological constraints on large extra dimensions," hep-ph/0101131.   

\bibitem{wells} G.~F.~Giudice, R.~Rattazzi and J.~D.~Wells, Nucl.\ Phys.\ {\bf B544}, 3 (1999) [hep-ph/9811291].

\bibitem{hewett} J.~L.~Hewett, Phys.\ Rev.\ Lett.\ {\bf 82}, 4765 (1999) [hep-ph/9811356].

\bibitem{peskin} E.~A.~Mirabelli, M.~Perelstein and M.~E.~Peskin, Phys.\ Rev.\ Lett.\ {\bf 82}, 2236 (1999) [hep-ph/9811337].

\bibitem{rizzo} T.~G.~Rizzo, Int.\ J.\ Mod.\ Phys.\ A {\bf 15}, 2405 (2000).

\bibitem{wheeler} C.~W.~Misner, K.~S.~Thorne and J.~A.~Wheeler, {\em Gravitation} (Freeman, San Francisco, 1973).

\bibitem{leblond2} F.~Leblond, R.~C.~Myers and D.~J.~Winters, "Consistency conditions for brane worlds of 
arbitrary codimension", in preparation.

\bibitem{appel} T.~Appelquist, A.~Chodos and P.~G.~Freund, {\em Modern Kaluza-Klein Theories} (Addison Wesley, USA, 1987);
E.~Witten, Nucl.\ Phys.\ {\bf B186}, 412 (1981). 

\bibitem{lykken} T.~Han, J.D.~Lykken and R.-J.~Zhang, Phys.\ Rev.\ D {\bf 59}, 105006 (1999) [hep-ph/9811350].

\bibitem{hyper} N.~Kaloper, J.~March-Russell, G.~D.~Starkman and M.~Trodden, Phys.\ Rev.\ Lett.\ {\bf 85}, 928 (2000)
[hep-ph/0002001].

\bibitem{lykken2} J.~Lykken and S.~Nandi, Phys.\ Lett.\ B {\bf 485}, 224 (2000) [hep-ph/9908505].

\bibitem{gegen} J.~S.~Gradshteyn and I.~M.~Ryzhik, {\em Table of Integrals Series and Products} (Academic Press, 
London, 1980).

\bibitem{bando} M.~Bando, T.~Kugo, T.~Noguchi and K.~Yoshioka, Phys.\ Rev.\ Lett.\ {\bf 83}, 3601 (1999) [hep-ph/9906549].

\bibitem{adel} C.~D.~Hoyle, U.~Schmidt, B.~R.~Heckel, E.~G.~Adelberger, J.~H.~Gundlach, D.~J.~Kapner and H.~E.~Swanson,
Phys.\ Rev.\ Lett.\ {\bf 86}, 1418 (2001) [hep-ph/0011014]. 

\bibitem{devi} A.~Kehagias and K.~Sfetsos, Phys.\ Lett.\ B {\bf 472}, 39 (2000) [hep-ph/9905417].

\bibitem{boas} R.~P.~Boas and C.~Stutz, Amer.\ J.\ Phys.\ {\bf 39}, 745 (1971).

\bibitem{leblond3} F.~Leblond, work in progress.

\end{thebibliography}
\end{document}